%% file: main.tex
\documentclass[sigconf, nonacm]{acmart} 

\usepackage{tikz}
\usepackage{caption}
\usepackage{subcaption}

\usepackage{afterpage}
\usepackage[export]{adjustbox}
\usepackage{blindtext}
\usepackage{subcaption}
\usepackage{graphicx}
\usepackage{lipsum}
\usepackage{xspace}
\usepackage{soul,color}
\usepackage{booktabs}
\usepackage{multirow}
\usepackage{array,boldline}
\usepackage{makecell}
\usepackage{algorithm}
\usepackage{algpseudocode}
\usepackage{float}
\usepackage{mathtools}
\usepackage{nicematrix}
\usepackage{url}
\hypersetup{
    colorlinks=true,
    linkcolor=blue,
    filecolor=magenta,      
    urlcolor=cyan,
}
\usepackage[framemethod=TikZ]{mdframed}
\AtBeginDocument{%
  \providecommand\BibTeX{{%
    \normalfont B\kern-0.5em{\scshape i\kern-0.25em b}\kern-0.8em\TeX}}}

\setcopyright{acmcopyright}
\copyrightyear{2021}
\acmYear{2021}

\acmConference[2021]
\acmBooktitle{XX '21: ACM - To Be Determined}

\begin{document}

\title{Towards Web Phishing Detection Limitations and Mitigation}


\author{Alsharif Abuadbba}
\email{sharif.abuadbba@data61.csiro.au}
\affiliation{%
  \institution{CSIRO's Data61 \& Cybersecurity CRC, Australia}
}

\author{Shuo Wang}
\email{Shuo.Wang@data61.csiro.au}
\affiliation{%
  \institution{CSIRO's Data61 \& Cybersecurity CRC, Australia}
}

\author{Mahathir Almashor}
\email{Mahathir.Almashor@data61.csiro.au}
\affiliation{%
  \institution{CSIRO's Data61 \& Cybersecurity CRC, Australia}
}
\author{Muhammed Ejaz Ahmed}
\email{Ejaz.Ahmed@data61.csiro.au}
\affiliation{%
  \institution{CSIRO's Data61, Australia}
}

\author{Raj Gaire}
\email{Raj.Gaire@data61.csiro.au}
\affiliation{%
  \institution{CSIRO's Data61 \& Cybersecurity CRC, Australia}
}

\author{Seyit Camtepe}
\email{Seyit.Camtepe@data61.csiro.au}
\affiliation{%
  \institution{CSIRO's Data61, Australia}
}
\author{Surya Nepal}
\email{Surya.Nepal@data61.csiro.au}
\affiliation{%
  \institution{CSIRO's Data61 \& Cybersecurity CRC, Australia}
}


\newcommand{\ballnumber}[1]{\tikz[baseline=(myanchor.base)] \node[circle,fill=.,inner sep=1pt] (myanchor){\color{-.}\bfseries\footnotesize #1};}

\renewcommand{\shortauthors}{Abuadbba, et al.}
\newcommand{\sharif}[1]{\textcolor{blue}{[Sharif: #1]}}

\input{Sections/0_Abstract}

\maketitle


\input{Sections/1_Introduction}

\input{Sections/2_Background}

\input{Sections/3_Insights}

\input{Sections/4_System_design}

\input{Sections/5_Evaluation}

\input{Sections/6_Discussion}

\input{Sections/7_Related_work}

\input{Sections/8_Conclusion}

\bibliographystyle{unsrt}
\bibliography{references}

\end{document}

%% file: Sections/0_Abstract.tex
\begin{abstract}

Web phishing remains a serious cyber threat responsible for the overwhelming majority of data breaches. Machine Learning (ML)-based anti-phishing detectors are seen as an effective countermeasure, and are increasingly adopted by web-browsers and software products. However, with an average of 10K phishing links reported per hour by platforms such as PhishTank and VirusTotal (VT), the deficiencies of such ML-based solutions are laid bare.

We first explore how phishing sites are bypassing ML-based detection with a deep dive into 13K phishing pages targeting major brands such as Facebook and Paypal. Results show successful evasion is caused by:
(1) use of benign services such as \texttt{sites.google} or \texttt{vu.co} to obscure phishing URLs;
(2) high similarity between the HTML structures of phishing and benign pages;
(3) hiding the ultimate phishing content within Javascript and running such scripts only on the client;
(4) looking beyond typical credentials and credit cards for new content such as IDs and documents;
(5) hiding phishing content until after human interaction.
We attribute the root cause to the dependency of ML-based models on the \textit{vertical feature space} (e.g., URL and webpage content). That is, these solutions rely \textit{only} on what phishers present within the page itself.

Thus, we propose Anti-SubtlePhish, a more resilient model based on logistic regression. The key augmentation is the inclusion of a \textit{horizontal} feature space, which examines correlation variables between the final render of suspicious pages against what trusted services have recorded (e.g., PageRank and WHOIS age). To harden against (1) and (2), we correlate information between WHOIS, Google index ranks, and page analytics. To combat (3), (4) and (5), we correlate features after rendering the page. Experiments with a mixture of 100K phishing and benign sites show promising accuracy for both reported and undetected phishing pages (96.1\% and 98.8\% respectively). Moreover, we obtained 100\% accuracy against 0-day phishing pages that were manually crafted, comparing well to the 0\% recorded by VT vendors over the first four days.

\end{abstract}

%% file: Sections/1_Introduction.tex
\section{Introduction}\label{sec:intro} 
Web phishing is a persistent cyber threat and responsible for over 90\% of data breaches~\cite{peng2019happens,kashapov2022email}. 
It tries to deceive a page visitor by pretending as a service the visitor knows in order to steal personal information such as login credentials. 
In addition, the proliferation of people online transactions nowadays such as bills, shopping and entertainment render the normal user exposure to phishing attacks inevitable. 
Phishing webpages costs American business half a billion dollars a year~\cite{mathews2017phishing}. 
Many anti-phishing solutions have been proposed to detect phishing attacks.
Examples include \textit{blacklists}~\cite{oest2019phishfarm}, \textit{heuristic}~\cite{teraguchi2004client}, \textit{similarity}~\cite{fu2006detecting,rosiello2007layout}, \textit{third-party services}~\cite{jain2018two} and \textit{machine learning} (ML)-based~\cite{pan2006anomaly} approaches. 
However, the nature of these \textit{blacklists} explains it’s shortcomings in not being able to detect 0-day phishing attacks \cite{oest2019phishfarm}.  \textit{Similarity} and \textit{heuristic}-based models might be able to detect 0-day phishing attacks, but have scalability and accuracy limitations \cite{zhang2007cantina}. Relying only on \textit{third-party services} (e.g., search engines) may have a few limitations as identified recently by Rao et al.~\cite{rao2019jail} such as high false negative in the case of new benign sites.

On the other hand, ML-based models are the only stream that shows the ability to detect 0-day phishing attacks while being scalable and accurate. Hence, they have been widely explored~~\cite{zhang2007cantina,whittaker2010large,thomas2011design,xiang2011cantina+,corona2017deltaphish,marchal2017off}  and integrated into industrial browsers (e.g., Chrome and Edge \cite{microsoft}). ML-based models depend on the assumption that phishing 
and legitimate pages contain statistically distinct patterns in the domains, HTML content or visual appearance~\cite{opara2019htmlphish}.
Consequently, ML-based techniques learn from an extracted set of features (e.g., URL characteristics, DOM structure, etc) from the content of phishing and benign websites, 
and later try to predict similar patterns. 
Hereby, the accuracy of these models relies heavily on the features selected  and how they are impervious to future phishing attacks. 

While the use of feature engineering techniques by ML-based anti-phishing tools have exhibited promises to defeat web phishing and have been adopted by many web browsers and software products, the number of reported phishing links every hour, 10,000 per hour on average, on the platforms like PhishTank~\cite{phishtank} and VirusTotal~\cite{virustotal} reveals a very different story. That is, these ML-based solutions are not as effective as they should be, which leaves significant questions (i.e., Why is that? How to improve?) yet to be addressed.
Therefore, this work is dedicated to first investigating the answers to the following  {\bf research questions (RQ):}

\vspace{0.2cm}
\begingroup
\leftskip=0cm plus 0.5fil \rightskip=0cm plus -0.5fil
\parfillskip=0cm plus 1fil
\noindent\textit{\textbf{ RQ1: Why ML-based approaches fail to capture recent phishing attacks?}}\par\endgroup
To answer RQ1, we first investigate the potential limitations 
by a deep dive case study across 13,000 phishing pages reported and manually verified to PhishTank in June 2020 that target famous victim brands including Facebook, PayPal, Google, Microsoft, eBay, and Amazon. 
Our results suggest five phishing trends for why ML-based on features extracted from only the webpage and its URL might be limited: 

\begin{enumerate}
    \item \textit{Use of benign web services to camouflage the phishing pages such as \texttt{sites.google}, \texttt{ddns.net}, \texttt{co.vu}.} For example, the attacker creates Facebook phishing page and deploy under benign service as (\textit{\underline{phishFB}.ddns.net}) which in turns bypass URL based phishing detectors such as~\cite{ma2009beyond,verma2015character,sahoo2017malicious, le2018urlnet, aung2019url} owing to \texttt{ddns.net} is a benign service.
    \item\textit{ High similarity in HTML structure between benign and phishing webpages due to the usage of benign services}. For instance, the attacker creates eBay phishing webpage and deploy under \texttt{sites.google}. This means the webpage skeleton structure of (\textit{\underline{phisheBay}.sites.google}) and (\textit{\underline{benign}.sites.google}) are barely distinguishable which in turns contributes to the detection failure of  ML-based anti-phishing  that rely on HTML distinguishable characteristics  such as \cite{li2019stacking, opara2019htmlphish,lei2020advanced}.
    \item \textit{Hiding the ultimate HTML content behind \textit{Javascript} to block feature extraction including the title/form and run only at the client browser}. \textit{Javascript} is a benign web programming language that allows the webpage owner to change the content at the user browser. The attackers exploit that to initially include benign content not related to phishing target (e.g., Facebook) then change the content only after bypassing the phishing detectors at the user browser when interacting with the webpage. 
    \item \textit{Looking for not only credentials and credit cards but also new custom content such as IDs and documents}. Many ML-based anti-phishing models have been developed to specifically aim to identify credentials related features such as in \cite{zhang2007cantina,xiang2011cantina+,tian2018needle,peng2019happens} which limit their efficacy when they faced with phishing webpages that target custom content such as IDs and documents. 
    \item \textit{Blocking the actual page content until human interacts with the page}. For example, the attacker  includes  \textit{reCAPTCHA} which is a benign tool that enables web hosts to distinguish between human and automated access to websites. In turns,  ML-based anti-phishing are faced with an empty incoming webpage which limits their ability to extract vital features. The webpage phishing content will only appear after the victim answers the \textit{reCAPTCHA} challenge.
\end{enumerate}

We derive the root cause as the dependency, to some extent,  on \textit{vertical feature-space} such as webpage URL and HTML structure. \textit{Vertical} refers to extracting in-depth features only from what attackers introduce within their  webpages. Motivated by these findings, we address the second RQ:

\vspace{0.2cm}
\begingroup
\leftskip=0cm plus 0.5fil \rightskip=0cm plus -0.5fil
\parfillskip=0cm plus 1fil
\noindent\textit{\textbf{ RQ2: How can we improve the recent ML-based approaches?}}\par\endgroup
\vspace{0.2cm}
We here focus on identifying potential features that  
not only rely on \textit{vertical feature space} which phishers present in their URL/ initial page content 
but also \textit{horizontal feature space} by correlating with collected information from multi-trusted services (e.g., WHOIS, PageRank and Selenium rendering). We find that it is possible to build a more resilient ML-based model from these \textit{horizontal feature space} with promising results against zero-day phishing attacks. This is due to \textit{the difficulty for the phishers to temper with not only the webpage itself (\underline{vertical feature space}) but also many other trusted services records} (\underline{\textit{horizontal feature space}}). 

Correspondingly, we have made the following contributions: 
\begin{itemize}
    \item We identify five new web phishing trends that consistently used to bypass existing ML-based web phishing detectors due to relying \textit{vertically} on features extracted from the webpage URL, HTML content and visual appearance.
    \item To mitigate these phishing attacks, we propose a novel feature extraction framework that not only relies \textit{vertically} on the initial page URL/content but also \textit{horizontally} by a correlation with collected information from multi-trusted services. Our collected features aim at capturing four identified components for robust classification: \textit{Page Reputation}, \textit{Goal}, \textit{Consistency} and \textit{Analytics}. 
    \item We develop a logistic regression-based model called Anti-SubtlePhish that relies on the identified vital features  to detect potential webpage phishing. 
    \item We extensively evaluate Anti-SubtlePhish against a large dataset of 100,000 benign/phishing webpages collected from PhishTank~\cite{phishtank}, OpenPhish~\cite{openphish} and Alexa~\cite{alexa}. We also evaluate against crafted zero-day phishing attacks simulating the above trends. Our obtained results demonstrate promising accuracy between 96.1\% and 98.8\% over the testing dataset and accuracy of 100\% over 0-day phishing attacks crafted dataset.
\end{itemize}
 

%% file: Sections/2_Background.tex
\section{Background}
This section provides the necessary information to understand our work. 
\subsection{Website and Webpage Structure}
\begin{figure}[!h]
\centering
\includegraphics[width=0.9\linewidth]{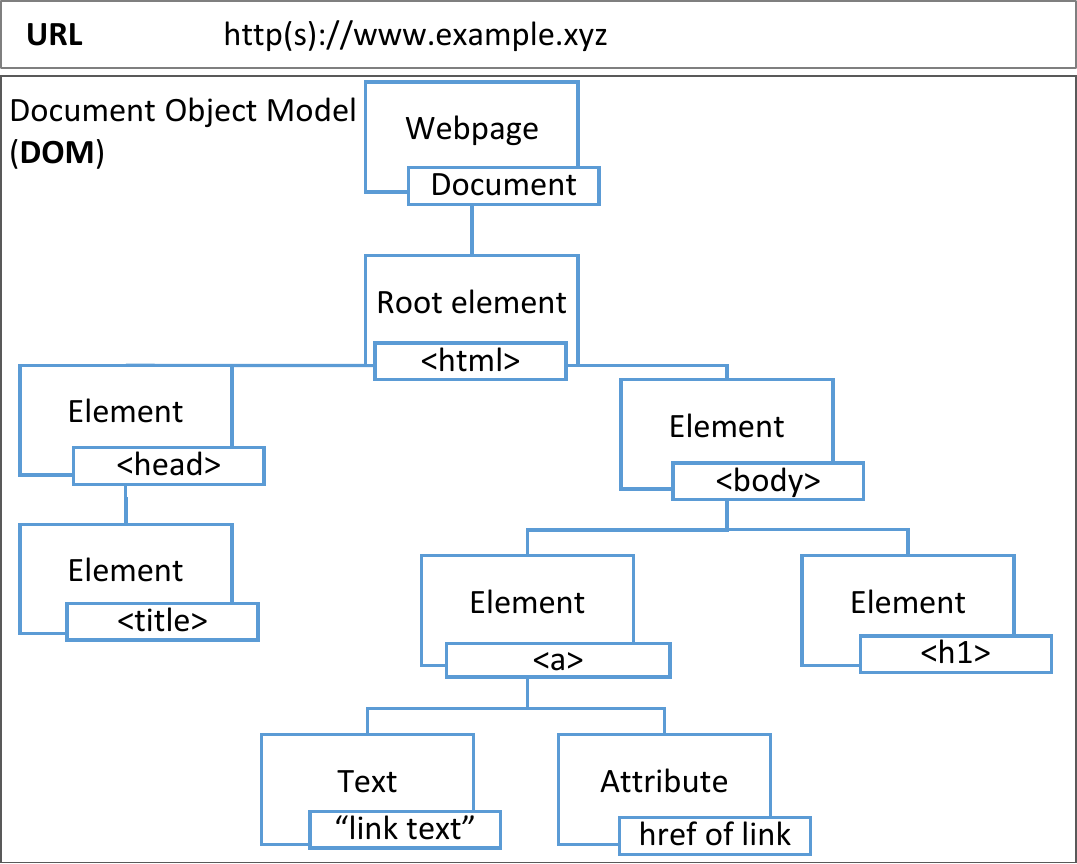}
\caption{Webpage Document Object Model Structure}
\label{fig:dom}
\end{figure}

A website is a collection of webpages that offer related content and can be identified by a common domain e.g., \textit{facebook.com}. 
Each webpage can be accessed via a unique URL such as \textit{www.facebook.com/login}. 
As depicted in Figure~\ref{fig:dom}, a webpage has a tree hierarchy called Document Object Model (DOM) which defines the logical structure of the page to be easily rendered and manipulated by the browsers. 
The tree has many types of nodes, named DOM nodes, including element (e.g., <a> hyperlink element), text (e.g., viewed link text), attributes (e.g., the actual link), and comment codes that are used only for illustration.

Besides, scripts such as \textit{Javascript} code are also a key part of the DOM tree which allows the webpage owner to manipulate its content at the user browser by \textit{adding}, \textit{modifying}, or \textit{deleting} the DOM tree structure. 
 \textit{Javascript} is a benign programming language used widely. However, the attackers  may also exploit it  to disguise their actual phishing until a later stage to run only after reaching the victim user as detailed in Section~\ref{sec:insights}.

\subsection{Web Phishing}
Web Phishing is a type of social engineering attack~\cite{chen2019gui} often used to steal user sensitive information, including login credentials and credit card details. 
It happens when a phisher firstly creates a fake webpage, masquerading as a trusted brand, and then distributes its URL to lure victim users to click on the link which in turn redirects them to that phishing webpage to extract users sensitive information \cite{peng2019happens}. 
Phishers also create fake websites to be as similar as possible to known entities' websites to the user in terms of visual appearance or/and URLs.

The life cycle of web phishing,  as extensively studied recently by Adam et al.~\cite{oest2020sunrise}, goes through various steps in a lifespan average of days or even hours. 
First, phishers start by configuring a fake website of a targeted victim brand before deploying publicly with a phishing URL.  
Then, the phisher launches a campaign to distribute the phishing URL through emails, instant messages, or text messages. Some victim users start visiting the fake website and few provide their sensitive information. 

\subsection{ML-based Anti-Phishing}
ML-based anti-phishing methods typically extract a set of features $x_1,x_2,..,x_n$ of webpage's URL, DOM trees, logos, icons, visual appearance, and elements of legitimate/phishing websites. 
These features are fed into learning algorithms along with the ground truth label $y$ (e.g., legit/phish)  which in turn produces a classification model to distinguish between legitimate and phishing websites. 
Based on many existing ML-based anti-phishing solutions such as~\cite{zhang2007cantina,whittaker2010large,thomas2011design,xiang2011cantina+,corona2017deltaphish,marchal2017off}, we can generalise the underline ML classifiers process into the following three stages:

\begin{itemize}
    \item \textbf{Hypothesis:} a model representation that tries to estimate or map input features $x_s$ to output $y$ using model parameters called $\theta$. $\theta$ starts with initial values while being updated during learning. The hypothesis can be presented as
    \begin{equation}
    h_\theta (x)=\theta_0+\theta_1 x_1+\theta_2 x_2+...+\theta_n x_n
    \end{equation}
    \item \textbf{Cost Function: } after calculating the hypothesis, the cost function is used to measure how far our estimation is from the ground truth $y$ and can be calculated for $m$ samples as 
    \begin{equation}
        J(\theta)=\frac{1}{2m}\sum_{i=1}^{m}h_\theta (x^{(i)})-y^{(i)}
    \end{equation}
    \item \textbf{Gradient Decent:} is a general function for minimizing the cost function to be as close as possible to the ground truth $y$. We use calculated cost J($\theta$), a learning rate $\alpha$ and partial derivative $\partial$ of the initial $\theta_i$ to calculate new parameter $\theta_j$ as  
    \begin{equation}
        \theta_j=\theta_j-\alpha \frac{\partial}{\partial\theta_i}J(\theta_i)
    \end{equation}
    These three stages will be repeated until the model produces the best convergence.
\end{itemize}

%% file: Sections/3_Insights.tex
\section{Key Insights:  New Webphishing trends Examples}\label{sec:insights}

The focus is to answer the first research question:

\vspace{0.2cm}
\begin{mdframed}[backgroundcolor=black!10,rightline=false,leftline=false,topline=false,bottomline=false,roundcorner=2mm]
\begingroup
\leftskip=0cm plus 0.5fil \rightskip=0cm plus -0.5fil
\parfillskip=0cm plus 1fil
\noindent\textit{\textbf{ RQ1:} Why ML-based approaches fail to capture recent phishing attacks?}\par\endgroup
\end{mdframed}

Phishing pages often aim to lure users by imitating commonly used legitimate sites that they already know such as Facebook or PayPal. 
Most of the existing ML-based web phishing detection methods start by extracting features from unrelated pages categorised as phishing and benign. Namely, the phishing pages might be collected in bulk from PhishTank and unrelated to benign pages that are  collected in bulk from another source like Amazon Top-sites links from Alexa.
While this process may produce reasonable detection accuracy, but nevertheless it is not taking the direct approach of inspecting benign sites against related phishing pages that target the same brand. Thus, it may fail to reveal some insights that assist us to understand potential trends and  limitations. 
Therefore, this motivates us to start by deep-dive case study, we call it \textit{the apple-vs-apple} approach to identify potential limitations. In this approach, we analyse the famous victim brands (e.g., Facebook) vs their phishing counterparts (Reported phishing page of Facebook).

\textbf{Our Insights Dataset.}  We build our own dataset  of 13000 phishing pages reported and manually verified to PhishTank~\cite{phishtank} in June 2020 that target \textit{top 6 famous victim brands including Facebook, PayPal, Google, Microsoft, eBay, and Amazon}. Those brands are selected as they have collectively over 4 billion users and understanding the phishing trends targeting them would have high significance. We then perform a deep dive case study across those 13000 phishing samples. In the following, we present five identified evasion strategies to limit the efficacy of existing ML-based detection mechanisms. 


\begin{figure}[!ht]
\centering
\includegraphics[width=0.8\linewidth]{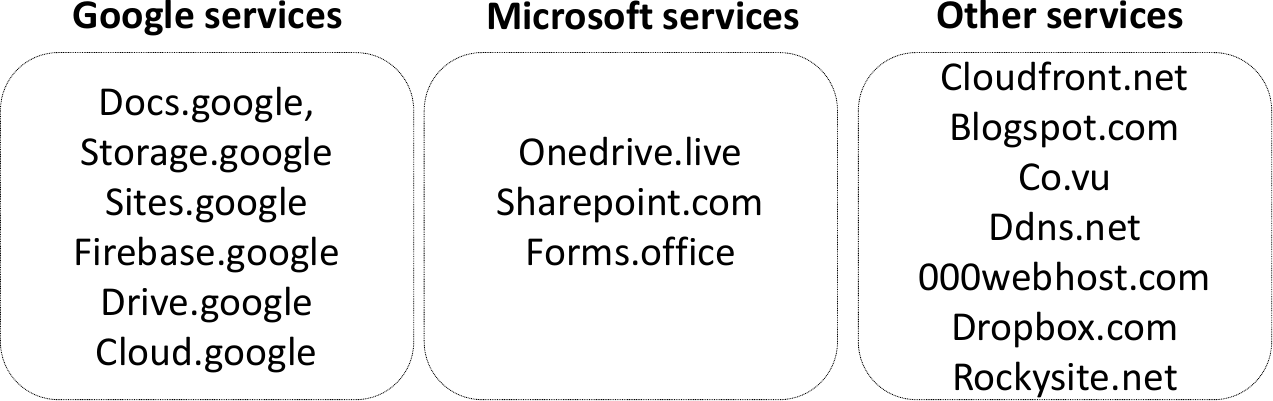}
\caption{Benign Services used to deploy phishing webpages.}
\label{fig:benignerviceList}
\end{figure}

\begin{figure*}[!ht]
\centering
\includegraphics[width=0.9\linewidth]{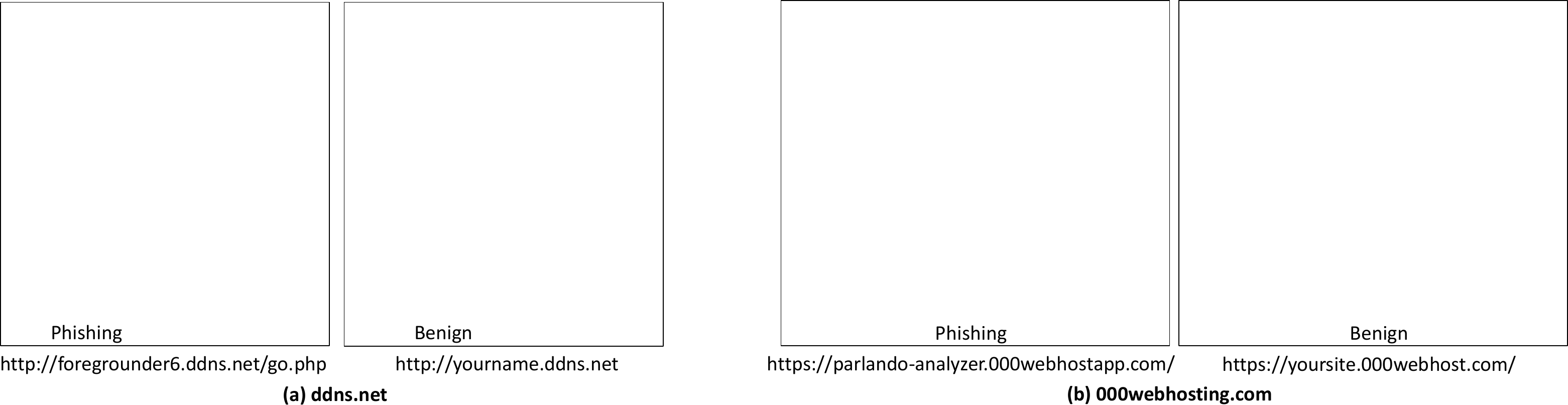}
\caption{Examples of the HTML DOM structure between phishing and benign pages on two benign services ddns.net and 000webhosting.com. The circles represents high level HTML tags such as <HTML> or <BODY>. The benign services used have an HTML template which demonstrates high similarity.}
\label{fig:htmlstructure}
\end{figure*}

\begin{figure}[!ht]
\centering
\includegraphics[width=1\linewidth]{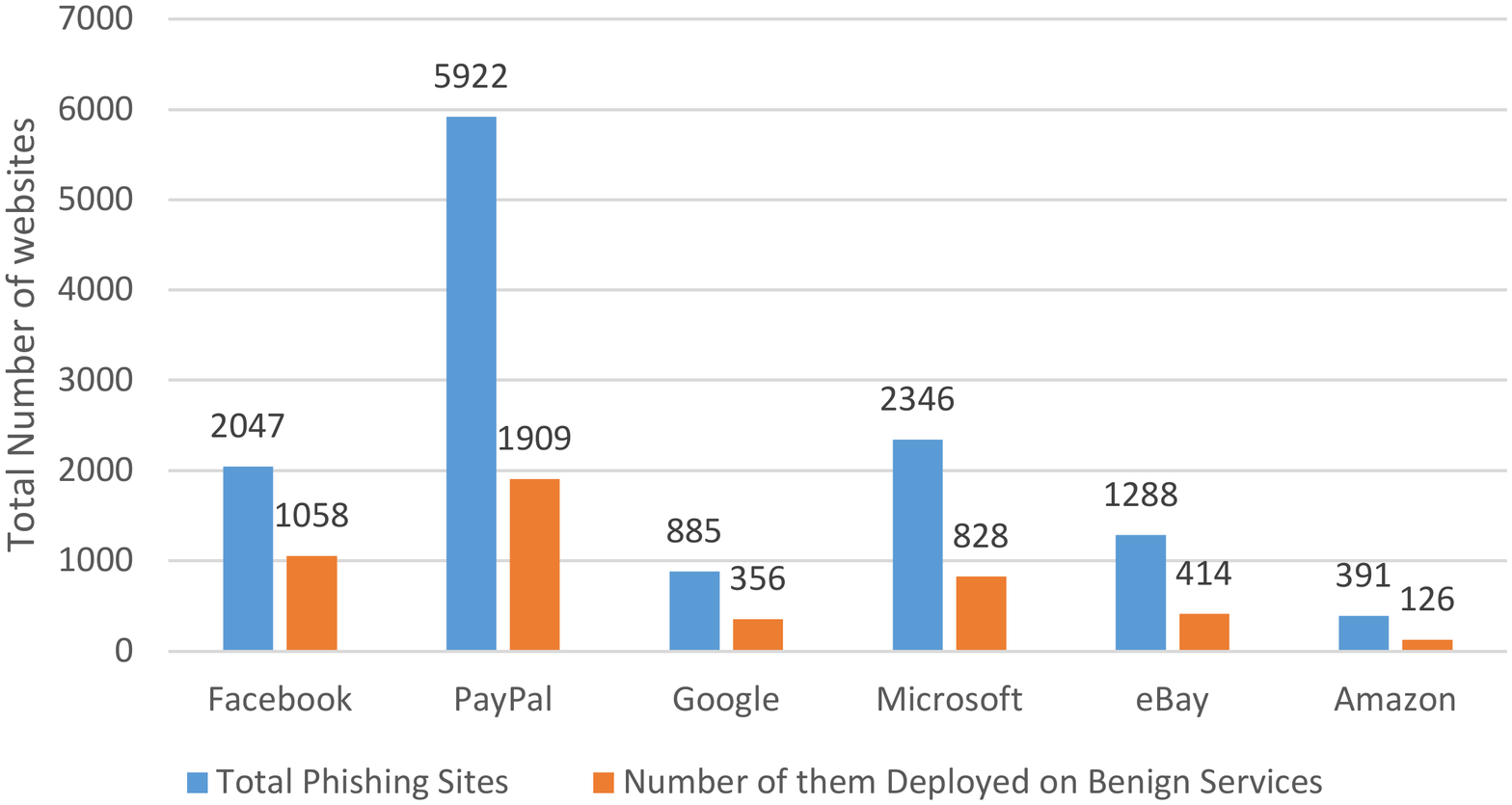}
\caption{Results of phishing sites that target the top 6 victim brands and reported to PhishTank in June 2020 against how many of these sites deployed on benign services to camouflage their URLs.}
\label{fig:victimon_benignervice}
\end{figure}
\subsection{Using Benign Web Services to Camouflage Phishing URLs}\label{subsec:benignserviceURL}
Many previous research studies have focused on the efficacy of phishing sites by analysing deceptive URLs (containing the name of the victim brand) v.s. random URLs. However, we are not aware of an existing study that looks beyond that to examine the proportion of where these sites are hosted. Our analysis suggests that there is a new phishing trend where attackers try to camouflage phishing URLs by deploying websites on benign services such as \texttt{sites.google},  \texttt{000webhostapp}, \texttt{ddns.net, co.vu}, etc. We identify 16 benign services as shown in Figure~\ref{fig:benignerviceList}. Recent concurrent work by De Silva et al.~\cite{de2021compromised} also confirmed a similar trend in more general malicious URL datasets collected from VirusTotal and not specific to phishing.  

Figure \ref{fig:victimon_benignervice} exemplifies our findings which shows a reasonable proportion of the phishing websites that target the top 6 victim brands are deployed on various benign web services. Namely, 2047 phishing websites target Facebook and 1058(51.6\%) of them are deployed on benign services.  5922 phishing websites targeting PayPal and 1909(32.2\%) are deployed on benign sites. 885, 2346, 1288 and 391 target Google, Microsoft, eBay, and  Amazon respectively. Out of these phishing sites, 32\%-40\% are deployed on benign services. Hence, while many ML-based anti-phishing models such as~\cite{ma2009beyond,verma2015character,sahoo2017malicious, le2018urlnet, aung2019url} that relies on detecting phishing by analysing the URLs may still effective to some extent, but nevertheless this trend limits their efficacy as the URL is from a benign service. In other words, the victim user is lured by a benign URL into a phishing webpage.

\subsection{High Similarity in HTML DOM Structure} 
Due to some of the limitations of the ML-based models relying only on the URL, many existing works \cite{li2019stacking, opara2019htmlphish,lei2020advanced} go beyond that using the HTML Document Object Model (DOM)\footnote{When a web page is loaded, the browser creates an HTML DOM of the page} structure to identify abnormal patterns as features to detect phishing. 
They do that by collecting two corpora of phishing and benign webpages.  
Although this stream of models may have some efficacy, but we identify a new trend that may limit their accuracy. That is, several phishing websites that target the top 6 brands and deployed on benign services (e.g., \textit{\underline{phisheBay}.ddns.net}) and have high HTML DOM structure similarity to other pages deployed on benign services (e.g., \textit{\underline{benign}.ddns.net}). 

Figure \ref{fig:htmlstructure} illustrates that by visualising two examples obtained from \texttt{ddns.net} and \texttt{000webhosting.com}. 
By using HTML2GDL\footnote{https://www.burlaca.com/2009/01/html2gdl/} library, we exemplify the visual  HTML DOM structure of benign and phishing websites deployed on two benign services. Table~\ref{tb:htmlstructure} shows  the percentages of phishing webpages that target the 6 top brands  and have their DOM structure with high similarity to benign services.
This indicates that attackers increasingly  employing this trend to ensure their phishing webpages have indistinguishable HTML DOM structure from benign services. 
Hence, the focus only on the HTML DOM structure as a source of ML-based features is less effective owing  to the difficulty of identifying any patterns between phishing and benign sites.

\begin{table}[!ht]
\centering
\caption{The percentages of phishing webpages with high DOM structure similarity to sites on benign Services.}
\scalebox{0.88}{
\begin{tabular}{ >{\centering\arraybackslash}m{0.4in} >{\centering\arraybackslash}m{0.3in} >{\centering\arraybackslash}m{0.3in} >{\centering\arraybackslash}m{0.5in} >{\centering\arraybackslash}m{0.3in} >{\centering\arraybackslash}m{0.3in}}\toprule

 $Facebook$ & $PayPal$ & $Google$ & $Microsoft$ & $eBay$ & $Amazon$\\ 
\hline
 24.4\% & 12.3\% & 23.9\% & 18.3\% & 16.4\% & 21.7\%\\

\hline

\bottomrule
\end{tabular}}
\label{tb:htmlstructure}
\end{table}
\subsection{Hiding the Ultimate HTML DOM Content behind Javascript} 
Usually, ML-based phishing detection systems scrape the HTML DOM structure of the incoming URLs and extract features  to classify webpages. 
During our analysis, we find a new trend that attackers recently used where they hide the final phishing page HTML structure behind a \textit{Javascript} call that only execute at the user browser.  
Web browsers have \textit{rendering} engine that takes HTML code and interprets it into what you see visually. It also call Javascript engine to execute any Javascript commands within the HTML and retrieve the relevant items from the remote servers like Images, which then manipulate the HTML elements. The Javascript part is the more dangerous because at the initial stage it looks as few commands but after rendering it could transform the content completely. 

\begin{figure}[!ht]
\centering
\includegraphics[width=0.9\linewidth]{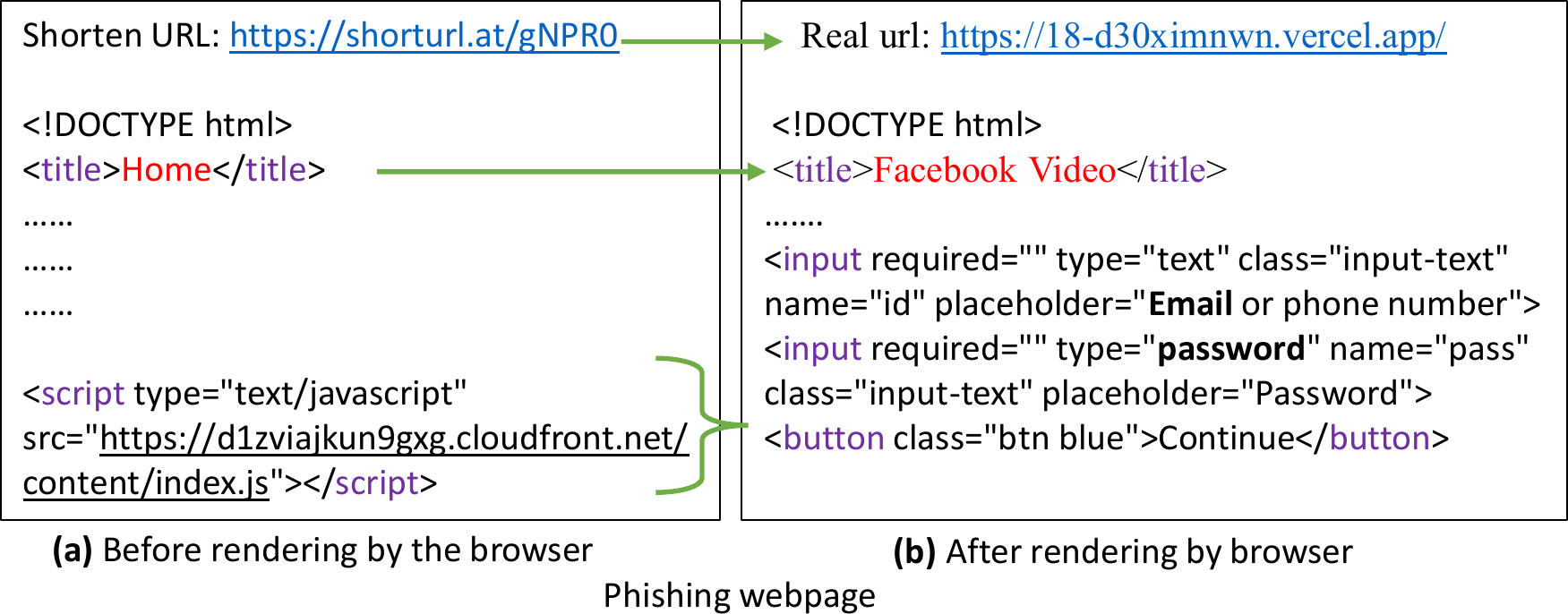}
\caption{Example of how phishers hide the final HTML DOM structure behind Javascript. (a) Prior rendering by the browser has nothing abnormal or related to Facebook. (b) Post rendering at the client browser where the phisher presents the actual goal to phish Facebook users}
\label{fig:javascript}
\end{figure}

Figure \ref{fig:javascript} shows an example from the phishing webpages we obtained from PhishTank on the 24th of June 2020. 
When we directly scrape the HTML DOM structure, we get the output (a) where the attacker uses shorten URL service, general title "\textit{Home}" and a Javascript call to "\textit{Cloudfront.net}" which is benign content (e.g., video, other files) delivery network offered by Amazon Web Services. 
However, after we render the webpage, using chrome rendering engine,  the Javascript call is executed and we get the output (b) where the URL is redirected to benign service to deploy front-end applications called "\textit{vercel.app}", the title turned into "\textit{Facebook Video}", and two input text-box appeared to collect email/password.
We detect 8\%-20\% of the phishing webpages using some form of Javascript camouflage to their actual intent as shown in Table~\ref{tb:javascript}. 
This trend indicates that the attackers try to bypass the existing ML-based detection system that relies on features extracted \textit{vertically} from only scraping the static webpage's URL and HTML DOM structure to build the classifier. Namely, \textit{the attackers conceal the malicious content including the title/forms and only rending them in late-stage  at the client browser which usually comes way after they bypass the classifiers.}

\begin{table}[!ht]
\centering
\caption{Phishing webpages using Javascript.}
\scalebox{0.88}{
\begin{tabular}{ >{\centering\arraybackslash}m{0.4in} >{\centering\arraybackslash}m{0.3in} >{\centering\arraybackslash}m{0.3in} >{\centering\arraybackslash}m{0.5in} >{\centering\arraybackslash}m{0.3in} >{\centering\arraybackslash}m{0.3in}}\toprule

 $Facebook$ & $PayPal$ & $Google$ & $Microsoft$ & $eBay$ & $Amazon$\\ 
\hline
 17.3\% & 9.4\% & 13.9\% & 19.8\% & 8.4\% & 16.7\%\\

\hline

\bottomrule
\end{tabular}}
\label{tb:javascript}
\end{table}

\begin{figure}[!ht]
\centering
\includegraphics[width=0.6\linewidth]{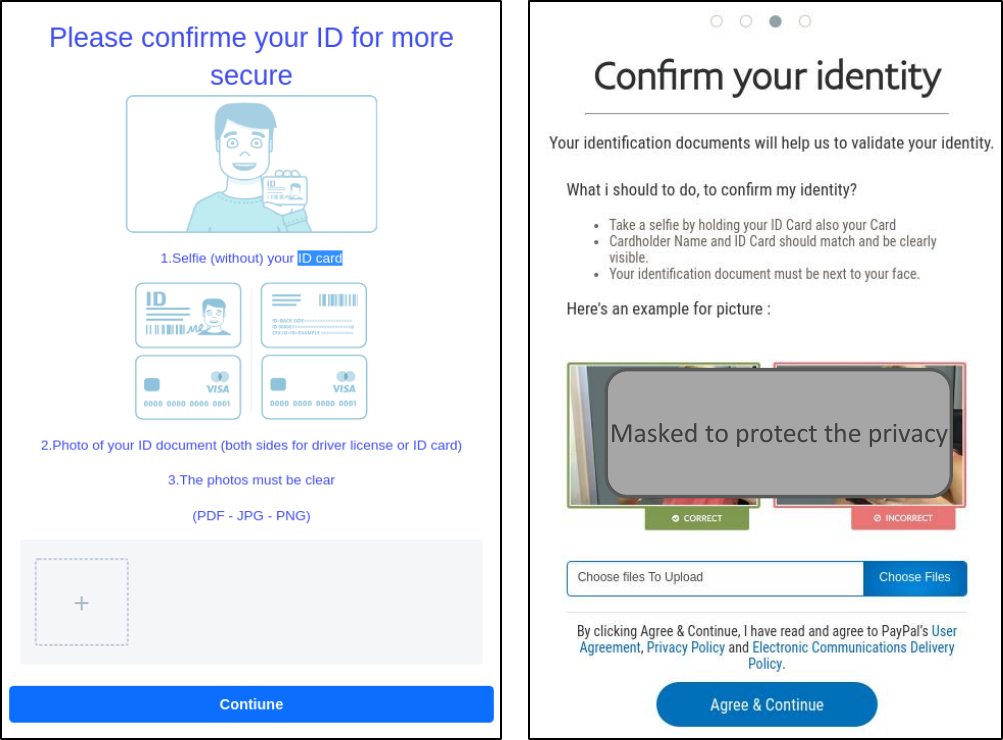}
\caption{Example of a new trend where phishers not only collect credentials but also target identity documents. They do that with industry-standard professionalism including illustrations and steps to guide the user through the process.}
\label{fig:identity}
\end{figure}
\begin{figure*}[!ht]
\centering
\includegraphics[width=0.8\linewidth]{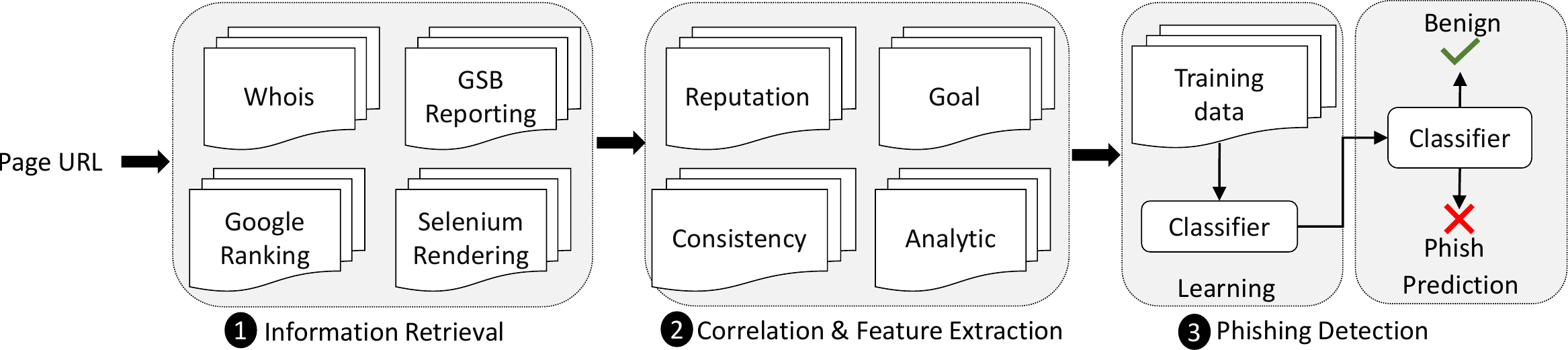}
\caption{High-level design of Anti-SubtlePhish}
\label{fig:design}
\end{figure*}
\subsection{Looking for not only Credentials but Broad Identity Documents}
Our analysis suggests that phishers are not only looking merely for credentials such as usernames and passwords, they also seek identity documents and photographs that they may monetize later. 
To be deceptive in luring the user, the attackers show a high level of professionalism similar to benign services by providing illustrations and steps to guide the user through as shown in Figure~\ref{fig:identity}. 
While this trend is less frequent than the previous three but still worth highlighting as it may limit the efficacy of  existing ML-based  models that developed intensively focusing on  detecting phishing  pages that look for credentials such as in \cite{zhang2007cantina,xiang2011cantina+,tian2018needle,peng2019happens}.

\subsection{Pretending as not Valid or Looking for Human Verification}
We observe that 5\%-15\% of phishing pages pretending to be not valid by presenting one of the known web interaction HTTP protocol tags such as ``\textit{Page Not Found 404}'', ``\textit{404 Error Page}'', ``\textit{OOPS! Page Not Exist}'', ``\textit{access Forbidden}'', etc. 
This is only presented within the body of the page but the actual HTTP protocol tag is "\textit{success 200}". In other words, the webpage is still active but only camouflaging the content due to phishing kit cloaking strategy which might be an effort to only target users from certain time zones or to avoid crawlers. 
Another rarely appeared but worth highlighting observation is that few pages presenting \textit{the reCAPTCHA} system with no other content, that tries to distinguish between human and automated access to websites. 
This is an effort from the phisher to phase out any web crawlers or ML-based systems that detect based on the page content and wait for human action to present the lure phishing content (e.g., login forms). 

\vspace{0.2cm}
\begin{mdframed}[backgroundcolor=black!10,rightline=false,leftline=false,topline=false,bottomline=false,roundcorner=2mm]

\noindent{\bf Summary:} To answer RQ1, we have identified 5 web phishing trends that may limit the efficacy of ML-based detection models which rely only on what phishers present in their web URL address or page content. 
\end{mdframed}

%% file: Sections/4_System_design.tex
\section{Anti-SubtlePhish System Design}
In this section, we provide the Anti-SubtlePhish detection system exploiting the above-identified insights to answer the RQ. 2:

\vspace{0.2cm}
\begin{mdframed}[backgroundcolor=black!10,rightline=false,leftline=false,topline=false,bottomline=false,roundcorner=2mm]
\begingroup
\leftskip=0cm plus 0.5fil \rightskip=0cm plus -0.5fil
\parfillskip=0cm plus 1fil
\noindent\textit{\textbf{ RQ2:} How can we improve the recent ML-based approaches?}\par\endgroup
\end{mdframed}

We first define the threat model that we focus on in this paper. We then provide an overview of the Anti-SubtlePhish detection system that can effectively distinguish phishing webpages from benign webpages.

\subsection{Threat Model}
We consider phishing pages targeting famous victim brands' websites as they have significant number of users. 
We assume that the attacker would be motivated to target websites that are widely known and trusted by using subtle phishing techniques such as camouflaging the URL, having high similarity in website HTML content with benign services (e.g., \textit{sites.google}) by deploying into these benign services, hiding the page goal behind a Javascript and running only at the user browser, or finally blocking the content until human interaction. 
We assume that the attacker could still craft the phishing page to be fully or partially similar to any page from the targeted websites.
We finally assume that the attacker sends those phishing webpages links through luring emails to victims. 

\subsection{Overview of Anti-SubtlePhish}
The overview of Anti-SubtlePhish is illustrated in Figure~\ref{fig:design}. 
It involves three stages: Information retrieval, correlation and feature extraction, and phishing detection. 
The aim of these stages is to overcome the challenges of relying on what phishers present initially to camouflage the page URL and HTML DOM structure as shown in Section~\ref{sec:insights}. 
\textit{The intuition is that we believe that multiple trusted domain can provide the information that help to ascertain the legitimacy of the webpage once correlated with the information within the webpage itself.}
Therefore, we retrieve information from multi-trusted services (e.g., \textit{WHOIS, Google Index, Selenium, GSB Reporting}) which we then correlate to extract reliable features that provide effective detection. 
While some of these services have been used solely to detect phishing, but nevertheless each has some limitations against newly registered legitimate sites and updated  URLs as studied by Rao et al.~\cite{rao2019phishdump}.
Therefore, by correlating the information from these trusted services together, we demonstrate the ability of better detection as will be shown shortly.

\subsubsection{\textbf{Information retrieval}} %
This is stage \ballnumber{1} as shown in Figure~\ref{fig:design} which  aims at collecting information from multi-trusted services. In specific, we  use the initial page URL to retrieve information from four services. 

\textbf{(a) WHOIS:} is a query and response protocol that is widely used for querying trusted databases that store some metadata about  domain names.  
We use Python wrapper (\url{https://pypi.org/project/whois/}) for Linux ``WHOIS'' command that is able to extract data for all the popular top-level domains such as  \textit{(com, org, net, biz, info, pl, jp, uk, nz, …)}.

\textbf{(b) Google Index:} is a search engine ranking results provided after rigorous examination from Google crawlers. 
We utilize Python wrapper (\url{https://pypi.org/project/google-api-python-client/} to hook up with Google Custom search  (\url{https://developers.google.com/custom-search/v1/overview}). 
We check the presence of domain names and retrieve the top ten matches with their metadata. 

\textbf{(c) Selenium Rendering:} is browser-based regression automation that enables simulation of rendering webpages as Chrome or Firefox. The reason behind that is to combat phishing trends where the goal of the page is hidden behind \textit{Javascripts}  and only run at the user browser. We use Python wrapper (\url{https://pypi.org/project/seleniumwrapper/}) to retrieve both the URL and the actual HTML DOM content before and after rendering. 

\textbf{(d) GSB Reporting:} after rendering the page using Selenium, we use Google safe browsing (GSB) reporting API to retrieve any available information about the domain name.  We use the publicly available API provided by Google (\url{https://developers.google.com/safe-browsing/}).

\subsubsection{\textbf{ Correlation \& Feature Extraction}} 

The stage \ballnumber{2}  aims at correlating and extracting potential features that would be reliable for the ML training and detection phase. We use the Pearson Correlation Coefficient~\cite{benesty2009pearson} to measure how closely two sequences of extracted features and produce correlated coefficient which could be  a robust feature for classification. 
From the collected features in stage 1, we identify 13 features from both \textit{vertical feature-space} and correlated with \textit{horizontal feature-space} under 4 categories -- proposed to capture comprehensive views about the webpage. The four categories are the website \textit{Reputation}, \textit{Goal}, \textit{Consistency} and \textit{Analytic}. Next, We  explain these features and categories in detail. 

\begin{figure}
     \centering
     \begin{subfigure}[b]{0.3\textwidth}
         
         \includegraphics[width=\textwidth]{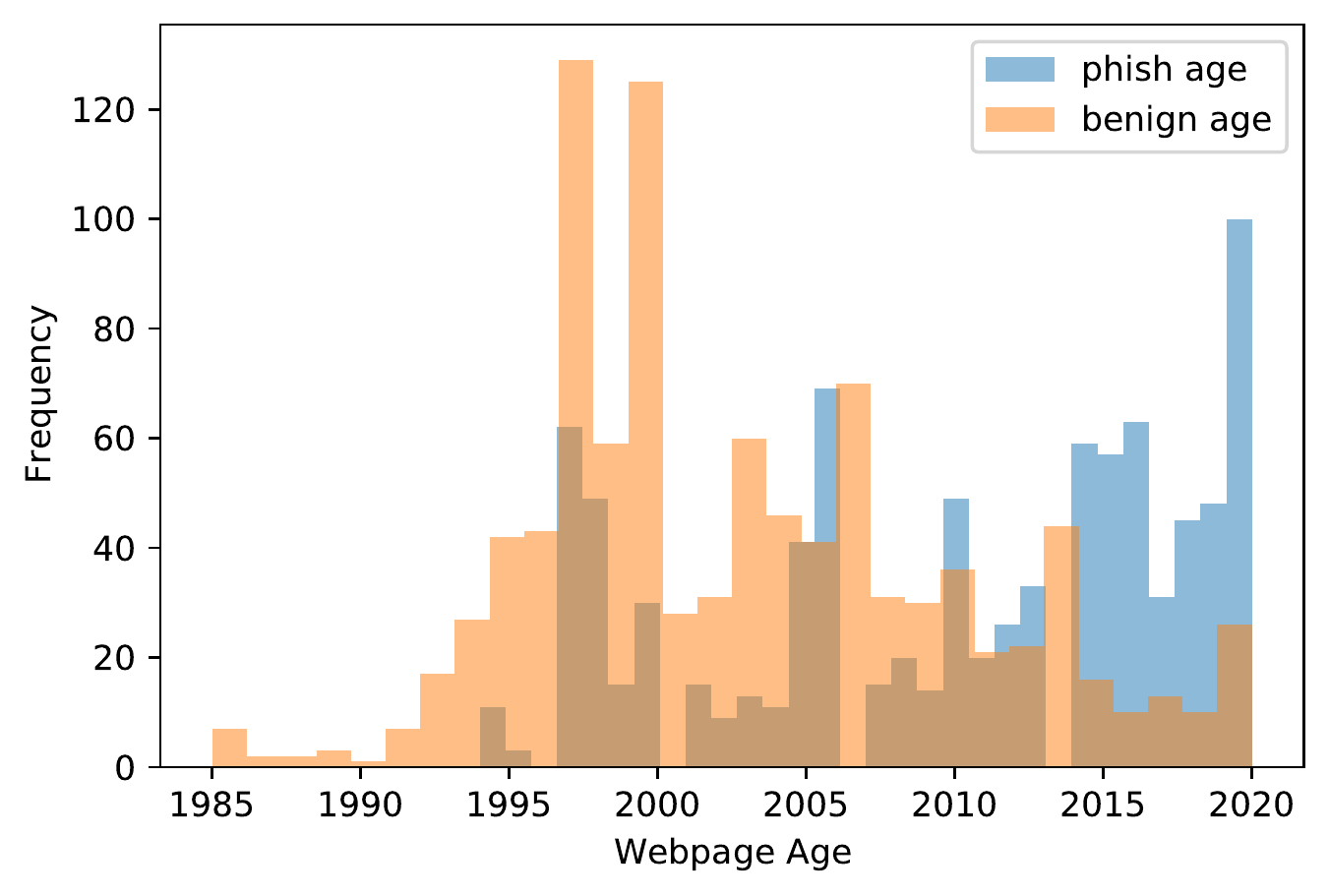}
         \caption{Histogram of the creation age retrieved from only WHOIS of a top 1000 benign sites vs 1000 phishing sites targeting them. The obtained values show wide overlap which means relying only on the WHOIS database might not be sufficient.}
         \label{fig:hist_age}
     \end{subfigure}
     \hfill
     \begin{subfigure}[b]{0.4\textwidth}
         \centering
         \includegraphics[width=\textwidth]{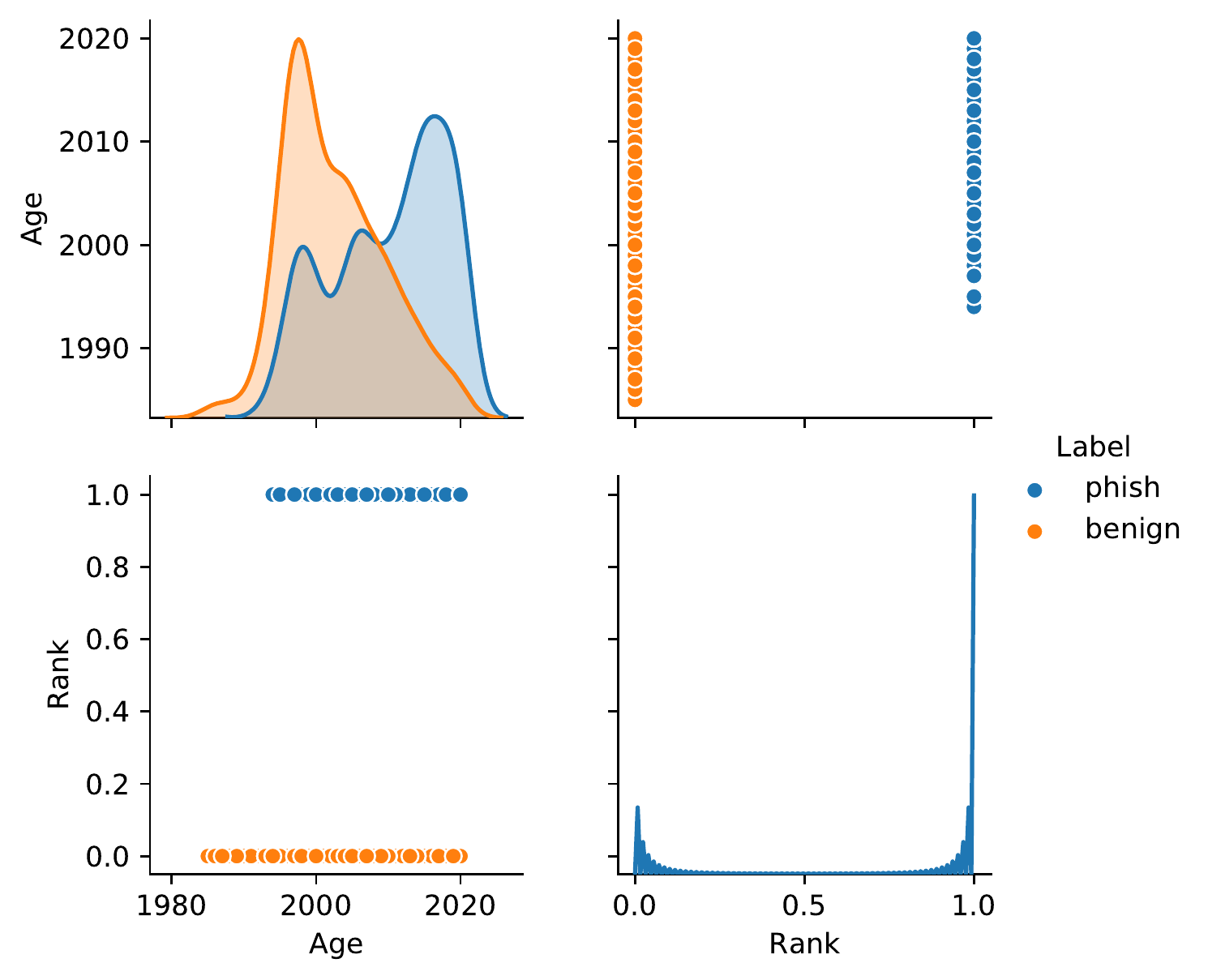}
         \caption{Correlating the website age with ranking demonstrates better reliable feature for ML model to classify benign vs phishing.}
         \label{fig:pair_age_rank}
     \end{subfigure}
        \caption{Example of correlating website creation age from WHOIS (a) with ranking from Google index as shown in (b) bottom-left would be preferable for ML models to detect.}
        \label{fig:three graphs}
\end{figure}

\textbf{(i) Reputation:} Under this category, we examine the web-page reputation in terms of validity, active duration(age), ranking, and reported by others as suspicious. 
For validity, we look for webpages that pretend to be not valid. To do so, we first render the webpage to execute all its scripts by using chrome rendering engine and Selenium. We then identify validity indicators using many selected keywords from exploring the phishing trends  such as "\textit{no longer available, not found, unpublished, does not exist, 404, access forbidden, etc.}". 
For the active duration(age), we identify that the public WHOIS service would provide \textit{Creation Date} as a tag field in their response. 
However, we experienced the challenge that registrars may use other tags for this field. We, therefore, identify many of these alternatives used by different registrars such as "\textit{Registration Time, Registered Date, Commencement Date, Changed Date, Registered On, Created On, etc.}". 
We also identify that relying only on the active duration(age) of the website with the assumption that phishing websites are always new would not be a reliable feature. 
Figure~\ref{fig:hist_age} shows the retrieved creation date of 1000 top sites vs 1000 phishing sites targeting them.
It is clear that they have a wide overlap which confirms our finding in the insight Section 3.1 as phishers are using benign services to deploy phishing sites. 
To overcome this challenge, we  correlate the active duration(age) with the obtained ranking of the page as shown in Figure~\ref{fig:pair_age_rank} bottom-left which clearly could be a reliable feature for detection.

\begin{figure}[!h]
\centering
\includegraphics[width=0.8\linewidth]{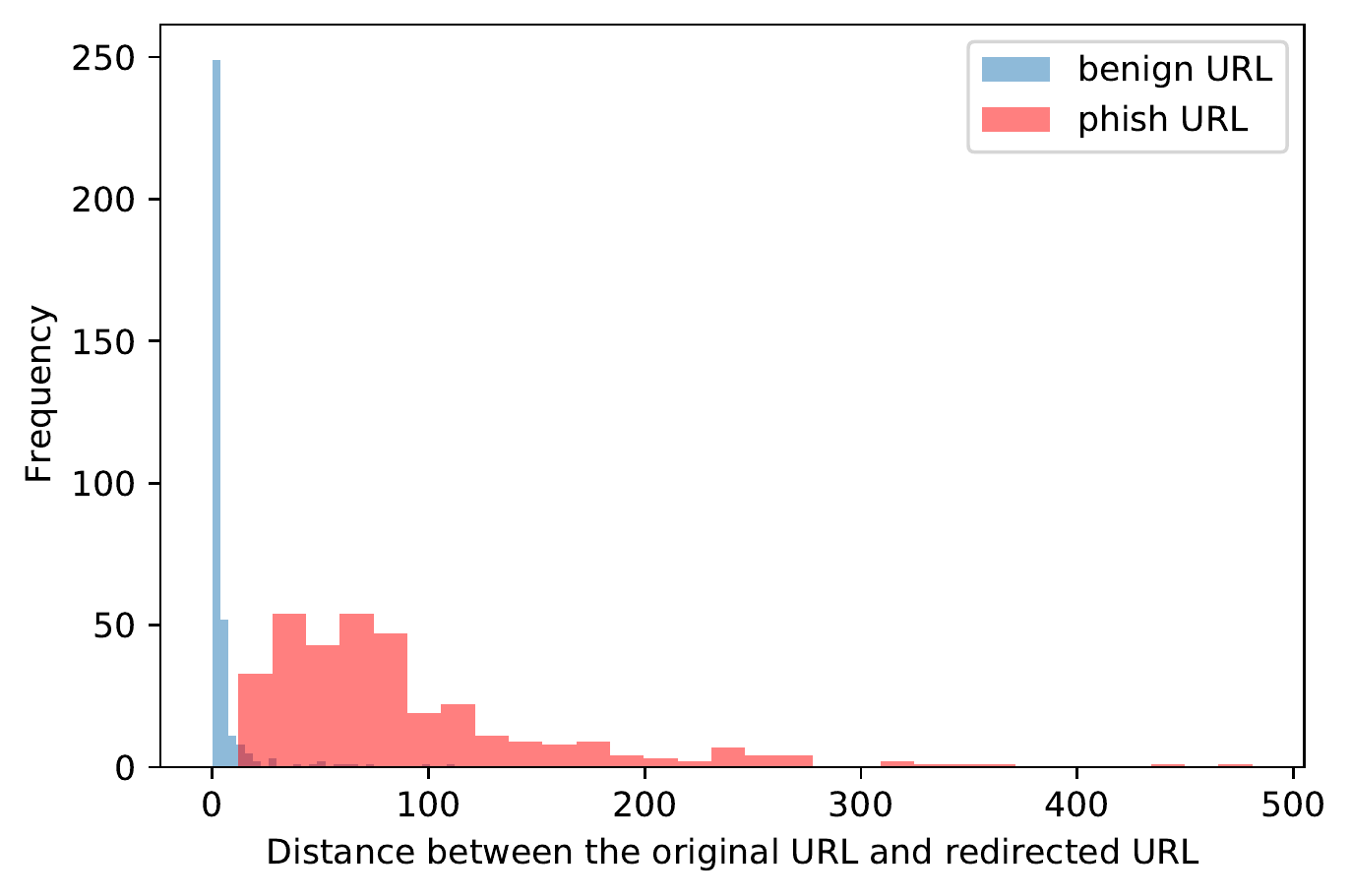}
\caption{Histogram of the distance between  the initial URL vs the actual URL after Selenium rendering. We see that the URLs distance of benign sites is close to zero, but it is much larger >30 for phishing websites.}
\label{fig:urldistance}
\end{figure}

\textbf{(ii) Goal:} Under this category, we concentrate on the webpage goal by examining two things: \textit{Does the webpage looks for input? Does the page redirect you? } 
To do so, we  correlate the initial URL with the actual URL after rendering the webpage using chromium engine and Selenium. We use Levenshtein distance\footnote{https://pypi.org/project/python-Levenshtein/} as a string metric for measuring the difference between two sequences. The Levenshtein distance $\nu$ between two strings $x,y$ (of length $|x|$ and $|y|$ respectively) is shown in Eq.~\ref{eq:Levenshtein}.
\begin{equation}\label{eq:Levenshtein}
 \nu(x,y)=
\left\{\begin{matrix}
|x| & if |x|=0,\\ 
|y| & if |y|=0,\\ 
\nu\left ( t(x),t(y) \right ) &if x[0]=b[0] \\ 
1+ min\left\{\begin{matrix}
\nu\left ( t(x),y \right ) & \\ 
\nu\left ( x,t(y) \right ) & otherwise,\\ 
\nu\left ( t(x),t(y) \right ) & 
\end{matrix}\right. & 
\end{matrix}\right.   
\end{equation}

where the $t$ of string $x$ is a sequence of all but the first character of $x$. Note the  elements in the minimum $min$ corresponds to rotations through both strings $x$ and $y$.
Figure~\ref{fig:urldistance} clearly demonstrates that while the distance between the initial URL vs the actual URL after the rendering is close to zero for benign sites, it is much larger, e.g., $>30$ for phishing websites. 
We also decide if the webpage is looking for input after rendering the page by identifying forms, credit card fields, passwords, emails, IDs, photographs, etc. 

\textbf{(iii) Consistency:} Under this category, we aim at the consistency of the initial URL, the actual URL, and the webpage HTML content after rendering. 
We also examine if the actual URL has been deployed on free hosting or benign services such as \texttt{sites.google, ddns.net, vu.co, branch.io, etc}

\textbf{(iv) Analytic:} Under this category, we  focus on performing analytics on actual URL-related features after rendering to identify common phishing trends. 
Among those, we selected a few that were already been identified by researchers and still applicable, such as IP address within the URL~\cite{zhang2007cantina}, suspicious symbols e.g., '\textit{@, -}'~\cite{gu2013efficient}, number of sub-domains~\cite{mohammad2014intelligent}, and domain length~\cite{mcgrath2008behind}. 
We also observe a new trend where even though the benign URL might be longer than phishing, but the main part of the domain of phishing websites is much longer than the benign in the case of setting up new domains.  

\subsubsection{\textbf{ Phishing Detection}} 
\begin{figure}[!h]
\centering
\includegraphics[width=0.8\linewidth]{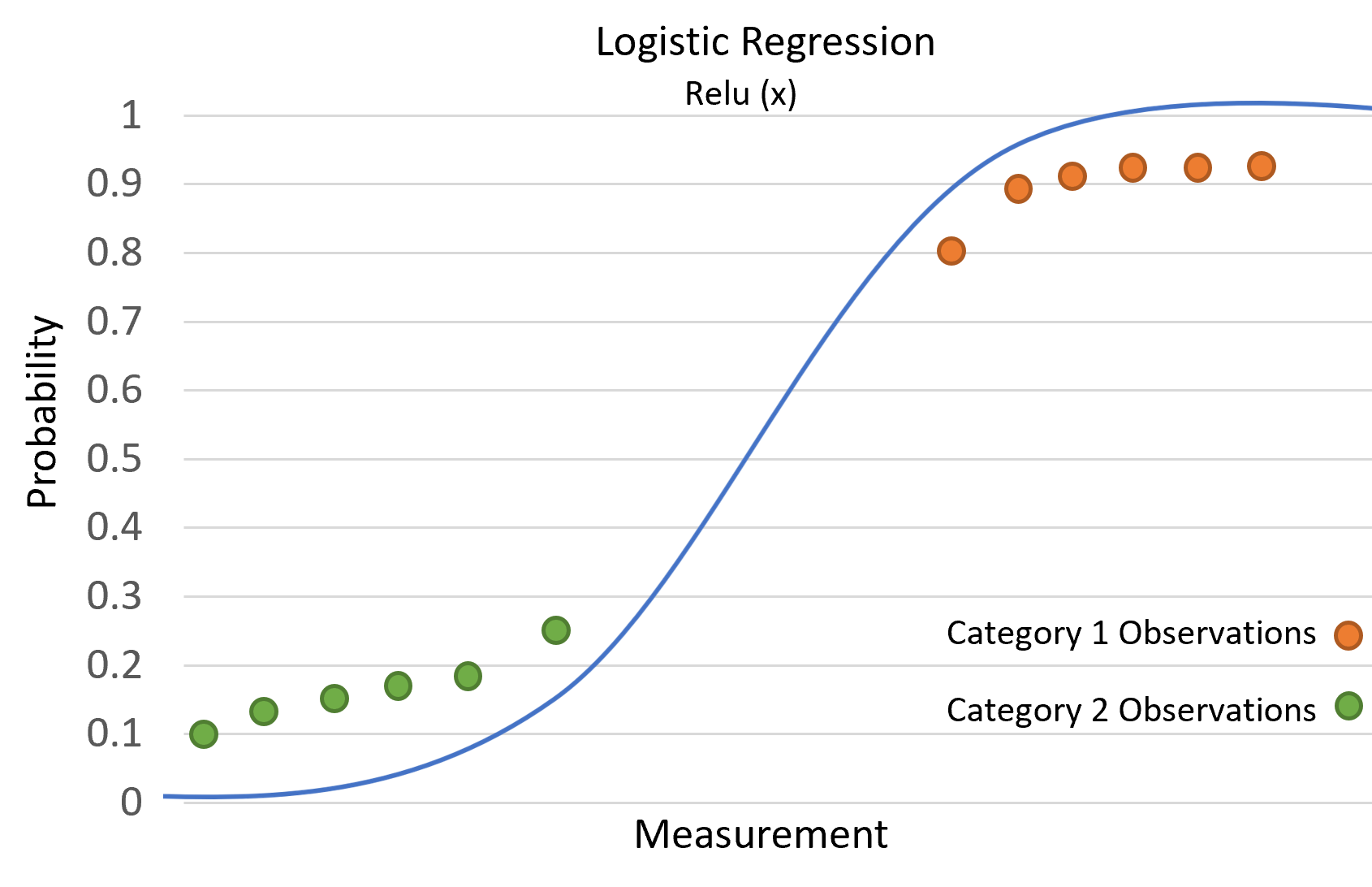}
\caption{Example of Logistic regression function ``relu'' that classifies input $X$ that has $n$ features into two district values as $0$ or $1$.}
\label{fig:lr}
\end{figure}

The stage \ballnumber{3} aims at training the ML model over the extracted features and measuring the efficacy. 
We select the Logistic Regression (LR) which is a common ML technique  to build a model that can discriminate between samples from two classes.  
The LR can be explained with Logistic function, also known as  \textit{Relu} function that take any real input $x$, and outputs a probability value between $0$ and $1$ which is defined in Eq.~\ref{eq:phishing_detection}.

\begin{equation}\label{eq:phishing_detection}
    h_\theta (x) = \frac{1}{1+e^{-\theta^Tx} }
\end{equation}
where $\theta$ is model parameters, $x$ is the input features and $T$ indicates transpose process.  
The learning process is similar to what we explained in Section 2.3 and the model fit using the above Logistic function can be visualised  as shown in Figure~\ref{fig:lr}.

\vspace{0.2cm}
\begin{mdframed}[backgroundcolor=black!10,rightline=false,leftline=false,topline=false,bottomline=false,roundcorner=2mm]

\noindent{\bf Summary:} To answer RQ2, we have developed a Logistic regression-based model that can learn from correlated features extracted from the information collected from multi-trusted services. We have demonstrated the efficacy of some of these correlations. Next, we focus on evaluating the detection accuracy of the developed model.
\end{mdframed}

%% file: Sections/5_Evaluation.tex
\section{Evaluation}
This section introduces the performance evaluation results of Anti-SubtlePhish including its efficacy against zero-day web phishing attacks in a comparison with industry-standard ML-based phishing detectors such as CyberCrime, Forcepoint, Fortinet, Kaspersky, Netcraft, and  Microsoft SmartScreen, and  VirusTotal other vendors\footnote{https://www.virustotal.com/gui/}.

\subsection{Dataset Collection and Experiments Settings}
To evaluate the performance of Anti-SubtlePhish, we collected benign and phishing pages from publicly available resources which is different from our insights dataset discussed in Section~\ref{sec:insights} to avoid bias. For phishing websites, we collected 50,000 URLs from PhishTank~\cite{phishtank} and OpenPhish~\cite{openphish}\footnote{https://openphish.com/} over the period Jun to September 2020. For benign website, we collected 50,000 URLs from Alexa Topsites Amazon service~\cite{alexa}\footnote{https://www.alexa.com/topsites}. We pass these URLs to the information retrieval stage \ballnumber{1} as shown in Figure~\ref{fig:design} to collect the relevant information from the aforementioned trusted entities. To evaluate a zero-day phishing attack scenario, we also setup 9 phishing sites using the identified insights. In summary, we eliminate any duplicates and split the dataset into four groups to carry out thorough experiments as follows:
\begin{itemize}
    \item Targeted dataset (\textbf{Dtarget}): Dtarget includes the top 1000 benign sites globally vs 1000 phishing sites that target top victim brands discussed in Section~\ref{sec:insights}.
    \item Small dataset (\textbf{Dsmall}): Dsmall includes the top 5000 benign sites vs 5000 random phishing sites to examine the efficacy of the obtained features.
    \item Large dataset (\textbf{Dlarge}): Dlarge includes 44,000 benign sites vs 44,000 phishing sites.
    \item Zero-day dataset (\textbf{Dzero}): Dzero includes 9 hand-crafted phishing sites that we set up to examine the potential of Anti-SubtlePhish in comparison to other algorithms against unseen phishing webpages using the identified insights.
\end{itemize}

\subsection{Evaluation Metrics}
The detection accuracy of Anti-SubtlePhish is evaluated with five metrics, accuracy, precision, recall, false acceptance rate (FAR), and false rejection rate (FRR), which are popularly used to evaluate the performance of classifiers.

\begin{itemize}
    \item \textbf{Accuracy\,(Acc.)} is the percentage of correctly classified webpages by a detection method.
    \item \textbf{Precision\,(Pre.)} is the percentage of webpages classified as phishing by a detection method, which are actual phishing webpages.
    \item \textbf{Recall\,(Rec.)} is the percentage of phishing webpages that were accurately classified by a detection method.
    \item \textbf{FAR} is the percentage of phishing webpages that are classified as benign webpages by a detection method.
    \item \textbf{FRR} is the percentage of benign webpages that are classified as phishing webpages by a detection method. 
\end{itemize} 

In general, while FRR is an indication of detection systems' reliability, FAR shows the security performance. Ideally, both FRR and FAR should be 0\%. Often, a detection system tries to minimize its FAR while maintaining an acceptable FRR as a trade-off, especially under security-critical applications. 

\subsection{Results of Web Phishing Detection }
Table \ref{tb:dtarget_results} shows the obtained results from our logistic regression-based Anti-SubtlePhish model vs various other ML models such as Decision Tree, KNN, Naive Bayes, Random Forest, and SVM. 
In this experiment, we use \textbf{\textit{Dtarget}} dataset. 
It is clear that our Anti-SubtlePhish could achieve the highest accuracy of 98.8\%, precision of 99.1\%, and recall of 97.8\% in comparison to other ML-based models. 
It also produces the lowest FAR of 0.9\% and FRR of 2.2\%. These results demonstrate that correlating collected features from multi-trusted entities could detect a substantial number of the subtle phishing tricks we identified against the top targeted sites.

\begin{table}[!ht]
\centering
\caption{Dtarget Dataset Results.}
\scalebox{0.9}{
\begin{tabular}{ >{\centering\arraybackslash}m{0.8in}  >{\centering\arraybackslash}m{0.3in} >{\centering\arraybackslash}m{0.3in} >{\centering\arraybackslash}m{0.3in} >{\centering\arraybackslash}m{0.3in} >{\centering\arraybackslash}m{0.3in} >{\centering\arraybackslash}m{0.3in}}\toprule

& $Acc.$ & $Prec.$ & $Rec.$ & $FAR$ & $FRR$\\ 
\hline
{$Logistic\; Reg.$} & 98.8\%	& 99.1\%	& 97.8\%	& 0.9\%	& 2.2\% \\ 

\hline

{$Decision\; Tree$} & 95.3\%	& 97.8\%	& 93.7\%	& 2.2\%	& 6.3\% \\
\hline
{$KNN$} & 93.7\%	&97.1\%	&90.9\%	&2.9\%	&9.9\% \\
\hline
{$Naive\; Bayes$} & 95.3\%	& 99.5\%	& 91.4\%	& 0.5\%	& 8.6\% \\
\hline
{$Random\; Forest$} & 95.4\%	&97.1\%	&94.9\%	&2.9\%	&5.9\%\\
\hline
{$SVM$} & 95.9\%	&98.3\%	&91.5\%	&1.7\%	&8.5\%\\
\bottomrule
\end{tabular}}

\label{tb:dtarget_results}
\end{table}

To generalise our findings to the non-targeted dataset, in the next experiment we use \textbf{\textit{Dsmall}} dataset that includes more diverse and general phishing pages. 
In specific, it includes 5000 benign sites and 5000 random phishing sites. 
Table \ref{tb:dsmall_results} shows the obtained results. 
While there is a slight drop in the accuracy compared to \textbf{\textit{Dtarget}} dataset, but nevertheless the Anti-SubtlePhish based logistic regression model still achieves high accuracy of 96.8\%, Precision of 97.7\%, Recall of 96.6\%. 
It also maintains a low FAR of 2.3\% and FRR of 3.4\%. This illustrates that correlating collected information from multi-trusted entities could also detect a diverse range of phishing sites.

\begin{table}[!ht]
\centering
\caption{Dsmall Dataset Results.}
\scalebox{0.9}{
\begin{tabular}{ >{\centering\arraybackslash}m{0.8in}  >{\centering\arraybackslash}m{0.3in} >{\centering\arraybackslash}m{0.3in} >{\centering\arraybackslash}m{0.3in} >{\centering\arraybackslash}m{0.3in} >{\centering\arraybackslash}m{0.3in} >{\centering\arraybackslash}m{0.3in}}\toprule

& $Acc.$ & $Prec.$ & $Rec.$ & $FAR$ & $FRR$\\ 
\hline
{$Logistic\; Reg.$} & 96.8\%	& 97.7\%	& 96.6\%	& 2.3\%	& 3.4\%\\ 

\hline

{$Decision\; Tree$} & 95.8\%	& 94.6\%	& 95.9\%	& 5.4\%	& 4.1\%\\
\hline
{$KNN$} & 94.9\%	& 93.8\%	& 93.6\%	& 6.2\%	& 6.9\%\\
\hline
{$Naive\; Bayes$} & 90.1\%	&95.1\%	&78.5\%	&4.9\%	&21.5\%\\
\hline
{$Random\; Forest$} & 96.1\%	&94.6\%	&96.2\%	&5.4\%	&3.8\%\\
\hline
{$SVM$} & 95.7\%	&92.3\%	& 96.5\%	& 7.7\%	& 3.5\%\\
\bottomrule
\end{tabular}}

\label{tb:dsmall_results}
\end{table}

\begin{figure*}[!t]
\centering
\includegraphics[width=1\linewidth]{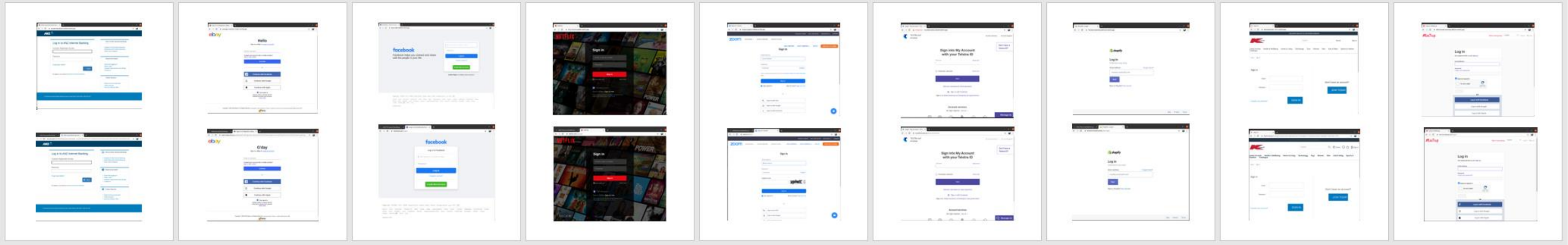}
\caption{Snapshots of the nine(9) phishing pages that we created for the zero-day attack experiments. Each white-page shows 2 images: (top) the original webpage, (bottom) the fake webpage we crafted.}
\label{fig:zeroday_snapshot}
\end{figure*}

To explore the Anti-SubtlePhish efficacy on a large scale general phishing dataset, in the next experiment we use \textbf{\textit{Dlarge}} that includes 44,000 benign sites and 44,000 phishing sites. 
Table \ref{tb:dlarge_results} presents the obtained results which demonstrate that our Anti-SubtlePhish based logistic regression model achieves high accuracy of 96.1\%, Precision of 96.7\%, Recall of 96.3\%. 
It also preserves a low FAR of 2.6\% and FRR of 3.6\% similar to the previous \textbf{\textit{ Dsmall}} experiments. It is also worth mentioning that in this large scale experiment, we have not removed any outliers from the phishing listed on PhishTank as many existing works do to be as close as possible to real-world scenarios. 
Hereby, our results indicate that correlating information from multi-trusted parties is a reliable way of detecting phishing websites.

\begin{table}[!ht]
\centering
\caption{Dlarge Dataset Results.}
\scalebox{0.9}{
\begin{tabular}{ >{\centering\arraybackslash}m{0.8in}  >{\centering\arraybackslash}m{0.3in} >{\centering\arraybackslash}m{0.3in} >{\centering\arraybackslash}m{0.3in} >{\centering\arraybackslash}m{0.3in} >{\centering\arraybackslash}m{0.3in} >{\centering\arraybackslash}m{0.3in}}\toprule

& $Acc.$ & $Prec.$ & $Rec.$ & $FAR$ & $FRR$\\ 
\hline
{$Logistic\; Reg.$} & 96.1\%	& 96.7\%	& 96.3\%	& 2.6\%	& 3.6\%\\ 

\hline

{$Decision\; Tree$} & 95.4\%	&95.6\%	&95.2\%	&3.4\%	&4.8\%\\
\hline
{$KNN$} & 91.6\%	& 94.3\%	& 88.4\%	& 5.7\%	& 11.6\%\\
\hline
{$Naive\; Bayes$} & 89.6\%	& 92.3\%	& 86.3\%	& 7.7\%	& 13.7\%\\
\hline
{$Random\; Forest$} & 94.3\%	& 93.8\%	& 93.1\%	& 6.2\%	& 6.9\%\\
\hline
{$SVM$} & 93.1\%	& 93.3\%	& 92.6\%	& 6.7\%	& 7.4\%\\
\bottomrule
\end{tabular}}

\label{tb:dlarge_results}
\end{table}
\subsection{Zero-day Phishing Detection}
One essential expectation from ML-based models is to detect new phishing pages called zero-day attacks.  
While our Anti-SubtlePhish based logistic regression model demonstrates reasonable detection efficacy, but nevertheless we previously tested only with already reported phishing pages to PhishTank. 
Therefore, to examine the efficacy of our Anti-SubtlePhish based model against zero-day web phishing attacks, we setup 9 phishing pages shown in Figure~\ref{fig:zeroday_snapshot}. The websites are \textit{Facebook, eBay, Netflix, Zoom, Shopify, ANZ, Telstra, Kmart, meetup}. We ensure selecting a wide range of services including  social media, e-shopping, banks, video conferencing, telecommunications, and streaming. We follow standard ethical considerations  including obtaining ethical clearance, reaching out to these organisations to let them know where possible and not capturing any credentials nor distributing any links. The steps of our experiments are as follows. (1) Setup zero-day phishing attack webpages of those brands with the  identified subtle phish trends highlighted in the \textit{Key Insights} Section \ref{sec:insights}.
(2) Deploy these 9 webpages under benign services with randomly chosen domain names (e.g.,  https://rU47H?urZBU8zBSd.co.vu) to avoid being inadvertently visited by real users.  
(3) Examine those 9 domain names against our Anti-SubtlePhish as well as various industry-standard ML-based phishing detectors via VirusTotal APIs such as CyberCrime, Forcepoint, Fortinet, Kaspersky, Netcraft, Microsoft SmartScreen, along with other VirusTotal  vendors.

\begin{figure}[!h]
\centering
\includegraphics[width=0.9\linewidth]{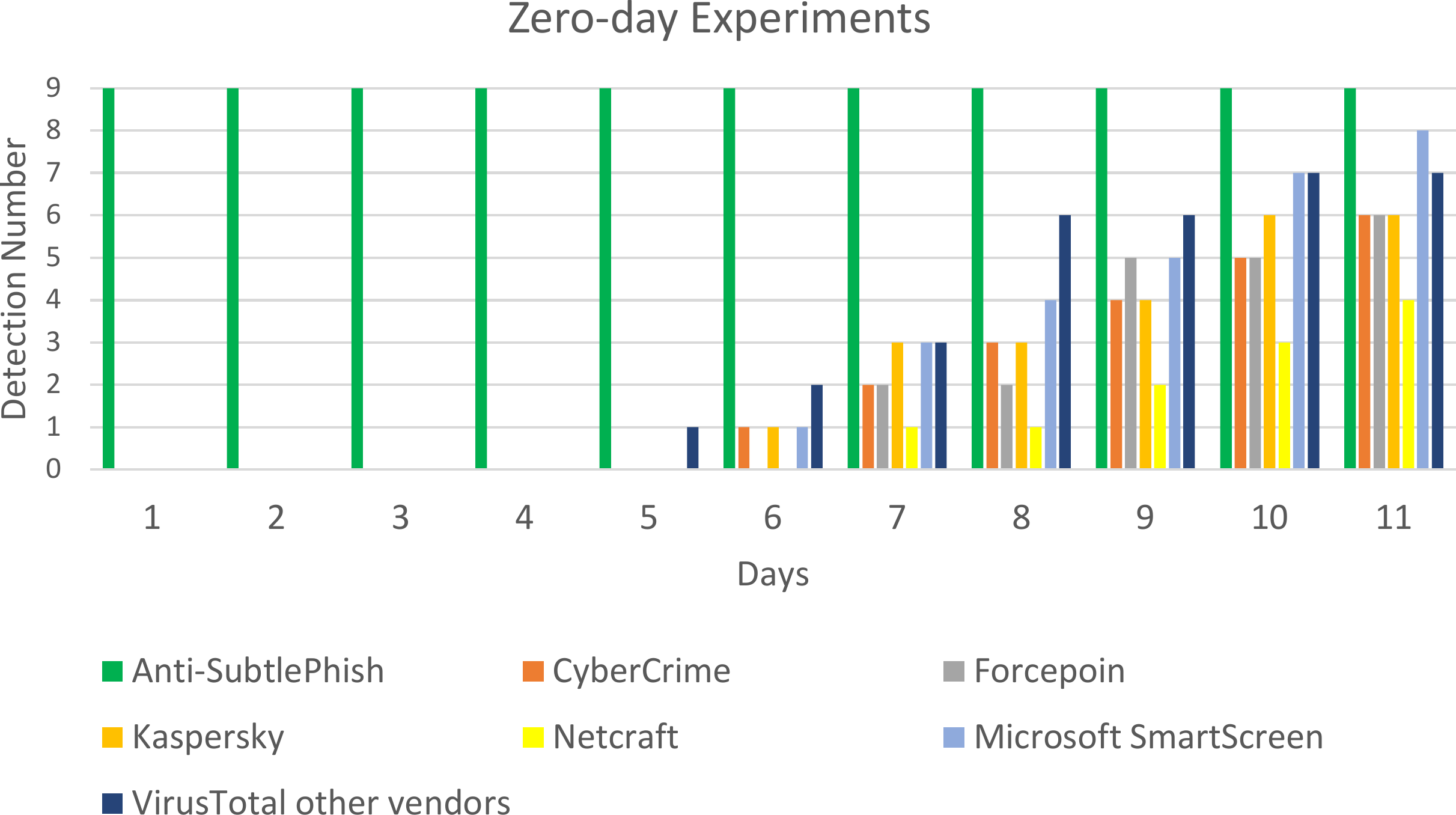}
\caption{Comparison between our Anti-SubtlePhish and various industrial ML-based detectors. Anti-SubtlePhish could detect all 9 zero-day crafted pages from the first day, whereas other vendors started to detect from the fifth day.}
\label{fig:zeroday_results}
\end{figure}

Figure~\ref{fig:zeroday_results} shows the obtained results over 11 days period. 
While our Anti-SubtlePhish detected all 9 out of 9 deployed pages on the first day, other industrial ML-based services did not flag these pages until the fifth day. Microsoft SmartScreen was able to flag 8 out 9 after 11 days. 
This illustrates the detection efficacy of our Anti-SubtlePhish based model by relying on correlating the collected information from multiple trusted services. 

%% file: Sections/6_Discussion.tex
\section{Discussion and Future Work}

\vspace{0.2cm}
\noindent{\bf Considerations for adaptive attacks:} Anti-SubtlePhish is built upon correlating collected information from not only the webpage itself but also multiple trusted parties such as WHOIS, Google Ranking and Selenium Rendering. We also noted from our key insights analysis that many existing adaptive attacks are developed to bypass a single source of information. For example, ML-based models ~\cite{ma2009beyond,verma2015character,sahoo2017malicious, le2018urlnet, aung2019url} that relies on using URLs as a single source for detection have been bypassed by deploying phishing webpages under benign hosting services. In other words, the content of the phishing webpage is still as it is but the URL is a benign link. A few other ML-based models \cite{li2019stacking, opara2019htmlphish,lei2020advanced} that rely on the DOM structure as a single source to identify phishing have been bypassed by using similar DOM structures to benign sites. Therefore, while it seems slightly easy for a phisher to develop an adaptive attack to bypass a single source of information, but nevertheless it would be more challenging to bypass not only the page information alone but also correlated information collected from multiple trusted entities.

\vspace{0.2cm}
\noindent{\bf Why detection accuracy dropped in the large scale experiment?}
We develop Anti-SubtlePhish from the insights obtained from \textbf{\textit{Dtarget}} that contains 1000 confirmed phishing webpages targeting the top famous brands. We achieved an accuracy of 98.8\%. On the other hand, we achieve an accuracy of 96.1\% in the large scale experiments. We use \textbf{\textit{Dlarge}} that contains 44,000 reported suspicious webpages to PhishTank where some of them might not be confirmed phishing yet. We also observe that the intention of a few of these webpages in \textbf{\textit{Dlarge}} might not be phishing to collect information but rather have popup screens and inappropriate adult content. We believe that some of these webpages contributed to a slightly lower accuracy compared to the \textbf{\textit{Dtarget}} dataset. Also, the comparable obtained results from our both datasets \textbf{\textit{Dsmall}} of 10K samples and accuracy at 96.8\% vs \textbf{\textit{Dlarge}} of 100K samples and accuracy at 96.1\% indicates that our model could produce stable accuracy.

\vspace{0.2cm}
\noindent{\bf Limitations and Challenges:}
Our Anti-SubtlePhish has the following limitations and challenges. (1) while \textit{WHOIS} information is available for most of the top-level domains (TLD) such as \textit{(com, org, net, biz, info, pl, jp, uk, nz, …)}, but nevertheless some of the TLDs might not be accessible through general \textit{WHOIS} which might be challenging especially in real-time deployment scenarios. One mitigation approach is to reach out to other regional Domain Name Service (DNS) servers to get information about these domains. (2) Some of the collected information from the trusted third party services   might not be organised nor easy to extract what we want autonomously. For example, \textit{WHOIS} information usually provides the age of the domain, but this piece of information might be presented under many tags such as "\textit{Registration Time, Registered Date, Commencement Date, Changed Date, Registered On, Created On, etc.}". During our experiments, we had to go through a large number of samples to identify those tags. However, we expect there might be  more variations of these tags which might be challenging on a real-time basis. One mitigation strategy might be to employ a more sophisticated unsupervised information extraction mechanism over unstructured data to be able to extract the dates from unseen tags.  (3) Most of the third-parties trusted entities may charge for the usage of collecting their information such as Google Ranking. They might also have imposed a daily limit (e.g., 10,000,  20,000, etc) on these services. Therefore, one would need to consider these challenges and obtain a custom access package to deploy Anti-SubtlePhish in real-time. 

%% file: Sections/7_Related_work.tex
\section{Related Work}
There are many anti-phishing solutions that have been proposed in the literature.
Examples include blacklists~\cite{oest2019phishfarm}, heuristic~\cite{teraguchi2004client}, similarity~\cite{fu2006detecting,rosiello2007layout} and machine learning (ML)-based~\cite{pan2006anomaly} approaches.  
Each category has some challenges. For instance,  the nature of these blacklists explains it’s shortcomings by not able to detect 0-day attacks \cite{oest2019phishfarm}. 
In addition,  similarity and heuristic-based models may be able to detect 0-day attacks but have scalability and accuracy limitations \cite{zhang2007cantina}. Lately, ML-based models are the only stream that shows the ability to detect 0-day attacks while being scalable and accurate. 
We next summarise the anti-phishing detection models based on the source of information they rely on to learn and detect.

\vspace{0.2cm}
\noindent{\bf Image Similarity:}  models under this category try to learn the similarity to existing legitimate websites. This similarity can be (1) \textit{Visual-based},  (2) \textit{Source-based}, or (3) \textit{Address-based}. 
In (1) \textit{Visual-based}, the model infers the similarity by learning from visual screenshots. 
As examples, Fu et al.~\cite{fu2006detecting} used Earth Mover's Distance (EMD) to compute similarity, then Zhang et al.~\cite{zhang2011textual} used EMD along with textual features. 
Cheng et al.~\cite{chang2013phishing}, Dunlop et al.~\cite{dunlop2010goldphish} used logo retrieval to determine a website identity that might be tricked by omitting the logo. 
Chen et al.~\cite{chen2010detecting} approximated human perception with Gestalt theory to decide the visual similarity. 
Recently, Sahar et al.~\cite{abdelnabi2020visualphishnet} introduced a triplet convolutional neural network to learn the similarity not only from the login page but rather the entire website. 
This stream  usually suffers from protecting only limited top famous brands in the magnitude of 100 to 150 pages. 
In (2) \textit{Source-based}, the similarity between phishing page to trusted page can be inferred by comparing the HTML content. 
Huang et al.~\cite{huang2010mitigate} extract the source content representation to compare against trusted identities. 
Liu et al.~\cite{liu2006antiphishing} segmented a webpage source content into blocks based on HTML visual cues to learn the  page similarity to trusted pages.  
This stream is vulnerable to code obfuscation as illustrated in~\cite{lam2009counteracting}. 
In (3) \textit{Address-based}, the similarity can be inferred by converting the page URL address to an image and training the model to later detect high similar suspicious URLs as demonstrated by Woodbridge et al.~\cite{woodbridge2018detecting}. 
This stream can be easily tricked by using random URL addresses. 

\vspace{0.2cm}
\noindent{\bf URLs:} models under this category tries to learn the representation of the legitimate URLs vs the phishing URLS to detect phishing~\cite{sahoo2017malicious,almashor2021characterizing}. 
Justin et al.  \cite{ma2009beyond} used statistical methods to discover the tell-tale lexical and host-based properties of malicious website URLs to predict the phishing sites. 
Rakesh et al. \cite{verma2015character} employed a two-sample Kolmogorov-Smirnov test along with other features extracted from the URLs and trained ML model to predict benign vs phishing URLs. 
Hung et al.~\cite{le2018urlnet} proposed a neural network model that learns from the character-level and word-level of the URLs built as a dictionary. 
Eint et al.~\cite{aung2019url} proposed a URL-based phishing detection model that learns  from the entropy of non-alphanumeric characters, which relies on the hypothesis that phishing  URLs differs from  legitimate ones  in their disorder structures. 
Despite the efficacy of models under this category, but they are prone to the adaptive attack of deploying the phishing webpage under benign service URLs, as shown in Section \ref{subsec:benignserviceURL}.

\vspace{0.2cm}
\noindent{\bf DOM structure:} models under this category try to go beyond the webpage URL and visual appearance into its HTML DOM structure to extract features that may help to distinguish phishing webpages from benign ones. 
Basnet et al.~\cite{basnet2011rule} detect by extracting features about alarm windows, hidden and restricted information within the DOM structure  as well as redirection patterns. 
Chidimma et al.~\cite{opara2019htmlphish} proposed a recurrent neural network-based model that learns from the entire DOM representation of benign and phishing webpages to be able to detect new unseen phishing webpages. 
Yukun et al.~\cite{li2019stacking} introduced a stacking model that combines 20 features from both the DOM  structure along with the URL to build an ensemble model for better detection. 
Rami et al.~\cite{mohammad2014intelligent} learned from the number of internal and external links within the DOM structure. 
Xiang et al.~\cite{xiang2011cantina+} detect only based on the information collection and forms. 
Alkhozae et al.~\cite{alkhozae2011phishing} rely on the hypothesis that the phishing webpages are relatively shorter than the legitimate pages and used that for detection. 
Despite the efficacy of some of these techniques,  various evasion mechanisms we presented in Section \ref{sec:insights} that highlights the limitations of relying only on DOM structure to detect phishing.
To name a few: (a) deploying phishing pages on benign services without changing the default DOM structure and (b) hiding the ultimate HTML DOM structure behind Javascript which is a widely adopted trend among benign websites as well. 

\vspace{0.2cm}
\noindent{\bf Third-party Service:} models under this category relies only on third-party services such as search engine ranking and Whois for detection. 
Mohammad et al.~\cite{mohammad2012assessment}, Rao et al.~\cite{rao2017enhanced} used age of the domain, registration date or expiry date to classify. 
Relying on these features only could be bypassed by deploying phishing webpages on benign services as we demonstrated in Figure~\ref{fig:pair_age_rank}. 
Chiew et al.~\cite{chiew2015utilisation,chiew2018leverage} used logo as a query to Google image search  to determine phishing and then improved that by using \textit{favicon} in search engine for detection.
Varshney et al.~\cite{varshney2016improving} extracted titles along with domains and query search engines to classify. 
Tan et al.~\cite{tan2016phishwho} used URL tokens and $N$ gram model for extracting the webpage identity keywords which are fed to search engines for detection. 
Jain et al.~\cite{jain2018two} introduced a two-level search engine based technique that relies on one domain/title as well as hyperlink based features for classification. 
Relying only on search engines may have a few limitations as identified recently by Rao and Pais~\cite{rao2019jail} such as high false negative in the case of deploying the phishing page into benign sites. 
On the other hand, they may lead to high false-positive in the case of newly registered legitimate sites. 

To overcome the highlighted shortcoming of the above categories, there is a compelled need to develop models that do not rely on a single source of information but rather correlating collected information from multiple trusted parties along with the rendering the webpage itself for better detection. 

%% file: Sections/8_Conclusion.tex
\section{Conclusion}
In this paper, we explored the potential reasons why many recent phishing sites bypass ML-based detection. 
This is achieved by  conducting a deep-dive case study across 13,000 phishing pages that target 6 top victim brands. 
We identified  five potential reasons for the  successful evasion. 
We derive the root cause as the dependency, to some extent,  on \textit{vertical feature-space} which limits the efficacy of the classifiers to be resilience against those types of evasion attacks.
To alleviate these attacks, we develop a logistic regression-based  model that relies not only on a single source of information comes from what phishers present in the webpage but also \textit{horizontal feature-space} by a correlation with third-party trusted  entities such as WHOIS, Google Index Ranking, Selenium rendering and page analytic. 
We devise a framework  of four elements for these features to evaluate page reputation, goal, consistency, and analytic. 
Our obtained results from 100,000 phishing/benign websites demonstrate promising accuracy of between 96.1\% and 98.8\%. as well as an accuracy of 100\% over 0-day attack crafted dataset. 
